\documentclass{elsart}                                               
\usepackage{amssymb}
\journal{Physics Letters B}
\begin{document}
\begin{frontmatter}

\title{Superconformal
Field Theory and SUSY N=1 KdV Hierarchy I:
Vertex Operators and Yang-Baxter Equation}
\author[POMI]{Petr P. Kulish}
\ead{kulish@pdmi.ras.ru}
\author[SpbU]{Anton M. Zeitlin}
\ead{zam@math.ipme.ru, http://www.ipme.ru/zam.html}
\address[POMI]{St. Petersburg Department of Steklov Mathematical 
Institute, Fontanka, 27,\\ St. Petersburg, 191023, Russia}
\address[SpbU]
  {Department of High Energy Physics, Physics Faculty, St. Petersburg State 
  University, Ul'yanovskaja 1, Petrodvoretz, St.Petersburg, 198904, Russia}
\date{20 May 2004}
\begin{abstract}
The supersymmetry invariant integrable structure of two-dimensional 
superconformal field 
theory is considered. The classical limit of the corresponding infinite family
of integrals of motion (IM) coincide with the 
family of IM of SUSY N=1 KdV hierarchy. The quantum version 
of the monodromy matrix, generating quantum IM, 
associated with the SUSY N=1 KdV 
is constructed via vertex operator representation of 
the quantum R-matrix. The possible applications to the perturbed 
superconformal models are discussed.   
\end{abstract}

\begin{keyword}
Superconformal field theory,
super-KdV,
Quantum superalgebras
\PACS 11.25.Hf; 11.30.Pb; 02.20.Uw; 02.20.Tw
\end{keyword}
\end{frontmatter}
\section{Introduction}
It is known that the superconformal field theory possesses two integrable 
structures \cite{gervais}. 
In the previous paper \cite{kulzeit} we have considered the quantum 
super-KdV \cite{sKdV1}-\cite{sKdV3} hierarchy,
which give rise to the infinite number of commuting integrals of motion (IM),
constructed via the generators of the superconformal algebra 
and therefore leading to one of these integrable structures.
But this set of IM is not invariant under 
the supersymmetry transformation.\\
\hspace*{5mm}Here we consider the SUSY N=1 KdV hierarchy \cite{mathieu},
\cite{inami} which is a supersymmetric generalization of the KdV one,
in the case of this model the supersymmetry generator is included in the
commuting family of IM \cite{gervais}.
The outline of the paper is as follows.
In the first part (Sec. 2) we consider the classical theory of SUSY N=1
KdV system, based on the twisted affine superalgebra $C(2)^{(2)}\simeq 
sl(1|2)^{(2)}\simeq osp(2|2)^{(2)}$. We introduce the supersymmetric
Miura transformation and monodromy matrix associated with the 
corresponding L-operator.
Then the auxiliary $\mathbf{L}$-matrices are constructed, 
which satisfy the quadratic Poisson bracket relation. As we will show in 
Sec. 4 quantum counterparts of these matrices coincide with a 
vertex-operator-represented quantum R-matrix.
The quantum version of the Miura transformation i.e. 
the free field representation of the superconformal algebra is given in 
Sec. 3. In Sec. 4 the quantum $C_{q}(2)^{(2)}$ superalgebra \cite{tolstkhor} 
is introduced.
Then it is shown that the corresponding quantum R-matrix can be represented
by two vertex operators, satisfying Serre relations of lower Borel subalgebra
of $C_{q}(2)^{(2)}$ and in the classical limit it coincides with 
$\mathbf{L}$-matrix.
The vertex-operator-represented quantum R-matrix $\mathbf{L}^{(q)}$
satisfies the so-called RTT-relation, which give us 
possibility to consider the 
model from a point of view of Quantum Inverse Scattering Method (QISM) 
\cite{leshouches}, \cite{kulsklyan}.
This could be well applied (Sec.5) to the study integrable 
perturbed superconformal theories with supersymmetry unbroken,
arising in the physics of  2D disordered systems, 
lattice models (e.g. the tricritical
Ising model),
and in the superstring physics (e.g. supersymmetric D-branes) 
\cite{qiu}-\cite{moore}.
\section{Integrable SUSY N=1 KdV hierarchy}
The SUSY N=1 KdV system can be constructed by means of the Drinfeld-Sokolov
reduction applied to the $C(2)^{(2)}$ twisted affine superalgebra 
\cite{inami}. 
The corresponding
$\mathcal{L}$-operator has the following form:
\begin{eqnarray}
\tilde{\mathcal{L}}_F=D_{u,\theta} 
-D_{u,\theta}\Phi (h_1+h_2)-\lambda({e_1}^{+}+{e_2}^{+}+{e_1}^{-}-{e_2}^{-}),
\end{eqnarray}
where $D_{u,\theta} =\partial_\theta + \theta \partial_u$ 
is a superderivative, the variable 
$u$ lies on a cylinder of circumference $2\pi$, $\theta$ 
is  a  Grassmann  variable, $\Phi(u,\theta)=\phi(u) - 
\frac{i} {\sqrt{2}}\theta\xi(u)$ 
is a bosonic superfield;
$h_1$, $h_2$, ${e_1}^{\pm}$, ${e_2}^{\pm}$ are the Chevalley 
generators of C(2) with the following commutation relations:
\begin{eqnarray} 
&&[h_1,h_2]=0,\quad [h_1,{e_2}^{\pm}]=\pm{e_2}^{\pm},
\quad [h_2,{e_1}^{\pm}]=\pm{e_1}^{\pm},\\
&&ad^{2}_{{e_1}^{\pm}}{e_2}^{\pm}=0, \quad ad^{2}_{{e_2}^{\pm}}{e_1}^{\pm}=0, 
\nonumber\\
&&[h_{\alpha},{e_{\alpha}}^{\pm}]=0 \quad(\alpha =1,2),\quad
[{e_{\beta}}^{\pm},{e_{\beta'}}^{\mp}]=\delta_{\beta, \beta'}h_{\beta}\quad
(\beta , \beta' =1,2),\nonumber
\end{eqnarray}
where the supercommutator [,] is defined as follows:
$[a,b]\equiv ad_{a} b\equiv ab- (-1)^{p(a)p(b)}ba$, where parity $p$
is equal to 1 for odd elements and is equal to 0 for even ones.
In the particular case of C(2) $h_{1,2}$ are even and $e^{\pm}_{1,2}$
are odd. 
The operator (1) can be considered as more general one, taken in the 
evaluation representation of $C(2)^{(2)}$: 
\begin{eqnarray}
\mathcal{L}_F=D_{u,\theta} 
-D_{u,\theta}\Phi h_{\alpha}-(e_{{\delta-\alpha}}+ e_{{\alpha}}),
\end{eqnarray}
where $h_{\alpha}$, $e_{{\delta-\alpha}}$, $e_{{\alpha}}$ are the 
Chevalley generators of $C(2)^{(2)}$ with such commutation 
relations: 
\begin{eqnarray}
&&[h_{\alpha_1},h_{\alpha_0}]=0,\quad [h_{\alpha_0},e_{\pm\alpha_1}]= 
\mp e_{\pm\alpha_1},\quad [h_{\alpha_1},e_{\pm\alpha_0}]= 
\mp e_{\pm\alpha_0},\\
&&[h_{\alpha_i},e_{\pm\alpha_i}]=\pm e_{\pm\alpha_i},\quad
[e_{\pm\alpha_i}, e_{\mp\alpha_j}]=\delta_{i,j}h_{\alpha_i},\quad (i,j=0,1),
\nonumber\\
&&ad^{3}_{e_{\pm\alpha_0}} e_{\pm\alpha_1}=0, \quad
ad^{3}_{e_{\pm\alpha_1}} e_{\pm\alpha_0}=0\nonumber
\end{eqnarray}
where $p(h_{\alpha_{0,1}})=0$, $p(e_{\pm \alpha_{0,1}})=1$ and 
$\alpha_1\equiv \alpha$, $\alpha_0\equiv\delta-\alpha$.
The Poisson brackets for the field $\Phi$, obtained by means of the 
Drinfeld-Sokolov
reduction are:
\begin{equation}
\{D_{u,\theta}\Phi(u,\theta), D_{u',\theta'}\Phi(u',\theta')\}=  
D_{u,\theta}(\delta(u-u')(\theta-\theta'))
\end{equation}
and the following boundary conditions are imposed on the components of 
$\Phi$: $\phi(u+2\pi)=\phi(u)+2\pi i p$, $\xi(u+2\pi)=\pm\xi(u)$.
The $\mathcal{L}_F$-operator is written in the Miura form, making a gauge
transformation one can obtain a new superfield $\mathcal{U}(u,\theta)\equiv
D_{u,\theta}\Phi(u,\theta)\partial_u\Phi(u,\theta)-D_{u,\theta}^3
\Phi(u,\theta)=-\theta U(u)-i\alpha(u)/\sqrt{2}$, 
where $U$ and $\alpha$ generate the superconformal algebra under the Poisson 
brackets:
\begin{eqnarray}
\{U(u),U(v)\}&=&
 \delta'''(u-v)+2U'(u)\delta(u-v)+4U(u)\delta'(u-v),\\
\{U(u),\alpha(v)\}&=&
 3\alpha(u)\delta'(u-v) + \alpha'(u)\delta(u-v),\nonumber\\
\{\alpha(u),\alpha(v)\}&=&
 2\delta''(u-v)+2U(u)\delta(u-v)\nonumber.
\end{eqnarray} 
Using one of the corresponding infinite family of IM, which are in involution
under the Poisson brackets 
(these IM could be extracted from the monodromy matrix 
of $\mathcal{L}_B$-operator, see below)\cite{gervais}:
\begin{eqnarray}
I^{(cl)}_1&=&\frac{1}{2\pi}\int U(u)\d u,\\
I^{(cl)}_3&=&\frac{1}{2\pi}\int
\Big(U^2(u)+\alpha(u)\alpha'(u)/2\Big)\d u,\nonumber\\
I^{(cl)}_5&=&\frac{1}{2\pi}\int
\Big(U^3(u)-(U')^2(u)/2-\alpha'(u)\alpha''(u)/4-
\alpha'(u)\alpha(u)U(u)\Big)\d u,\nonumber\\
& &   .\qquad.\qquad.\nonumber
\end{eqnarray}
one can obtain an evolution equation; for example, taking $I_2$
we get the SUSY N=1 KdV equation \cite{mathieu}:
$\mathcal{U}_t=-\mathcal{U}_{uuu}+3(\mathcal{U} D_{u,\theta}\mathcal{U})_u$ 
and in components:
$
U_t=-U_{uuu}-6UU_u - \frac{3}{2}\alpha\alpha_{uu}$, 
$\alpha_t=-4\alpha_{uuu}-3(U\alpha)_u.$
As we have noted in the introduction one can show that the 
IM are invariant under supersymmetry transformation generated by
$\int_0^{2\pi}\d u\alpha(u)$.\\
\hspace*{5mm}In order to construct the so-called monodromy 
matrix we introduce the 
$\mathcal{L}_B$-operator, equivalent to the $\mathcal{L}_F$ one:
\begin{eqnarray}
\mathcal{L}_B=\partial_u-\phi'(u)h_{\alpha_1}
+(e_{\alpha_1}+e_{\alpha_0}-\frac{i}{\sqrt{2}}\xi h_{\alpha_1})^2
\end{eqnarray}
The equivalence can be easily established if one considers the linear 
problem associated with the $\mathcal{L}_F$-operator: $\mathcal{L}_F\chi(u,
\theta)=0$ (we consider this operator acting in some representation of
$C(2)^{(2)}$ and $\chi(u,
\theta)$ is the vector in this representation). 
Then, expressing $\chi(u,\theta)$ in components:  
$\chi(u,\theta)$=$\chi_0(u)+\theta\chi_1(u)$, we find:
$\mathcal{L}_B\chi_0=0$ and $\chi_1=(e_{\alpha_1}+e_{\alpha_0}-\frac{i}{\sqrt{2}}\xi h_{\alpha_1})\chi_0$.\\
\hspace*{5mm} The solution to the equation $\mathcal{L}_B\chi_0=0$ can be written in the following way:
\begin{eqnarray}
\chi_0(u)&=&
e^{\phi(u)h_{\alpha_1}}P\exp\int_0^u \d u'\Big(\frac{i}{\sqrt{2}}
\xi(u')e^{-\phi(u')}e_{\alpha_1}\\
&-&\frac{i}{\sqrt{2}}
\xi(u')e^{\phi(u')}e_{\alpha_0}
-e^2_{\alpha_1}e^{-2\phi(u')}-
e^2_{\alpha_1}e^{2\phi(u')}-[e_{\alpha_1},e_{\alpha_0}]
\Big)\eta,\nonumber
\end{eqnarray}
where $\eta$ is a constant vector in the corresponding representation 
of $C(2)^{(2)}$.
Therefore we can define the monodromy matrix in the following way:
\begin{eqnarray}
\mathbf{M}&=&e^{2\pi i ph_{\alpha_1}}
P\exp\int_0^{2\pi} \d u\Big(\frac{i}{\sqrt{2}}
\xi(u)e^{-\phi(u)}e_{\alpha_1}\\
&-&\frac{i}{\sqrt{2}}
\xi(u)e^{\phi(u)}e_{\alpha_0}
-e^2_{\alpha_1}e^{-2\phi(u)}-
e^2_{\alpha_1}e^{2\phi(u)}-[e_{\alpha_1},e_{\alpha_0}]
\Big).\nonumber
\end{eqnarray}
Introducing then (as in \cite{1}) the auxiliary $\mathbf{L}$-operators:
$\mathbf{L}=e^{-\pi ip h_{\alpha_1}}\mathbf{M}$ we find that in the 
evaluation representation
(when $\lambda$, the spectral parameter appears) the following Poisson 
bracket relation is satisfied \cite{fadd}:
\begin{eqnarray}\label{eq:LLr}
\{\mathbf{L}(\lambda)\otimes_{,}\mathbf{L}(\mu)\}=
[\mathbf{r}(\lambda\mu^{-1}),\mathbf{L}(\lambda)\otimes \mathbf{L}(\mu)],
\end{eqnarray}
where $\mathbf{r}(\lambda\mu^{-1})$ is trigonometric $C(2)^{(2)}$ r-matrix 
\cite{shadrikov}.
From this relation one obtains that the 
supertraces of monodromy matrices 
$\mathbf{t}(\lambda)=str\mathbf{M}(\lambda)$ 
commute under the Poisson bracket:
$\{\mathbf{t}(\lambda),\mathbf{t}(\mu)\}=0$.
Expanding $\log(\mathbf{t}(\lambda))$ in 
$\lambda$ in the evaluation representation corresponding
to the defining 3-dimensional 
representation of $C(2)$ $\pi_{1/2}$ we find:
\begin{eqnarray}
\lim_{\lambda\to \infty}\log(\mathbf{t}_{1/2}(\lambda))=\sum^{\infty}_{n=1}c_n 
I^{(cl)}_{2n-1}{\lambda}^{-4n+2},
\end{eqnarray}
where $c_n=(-1)^{n-1}\frac{2^n}{n!}(2n-1)!!$. 
So, one can obtain the IM from the
supertrace of the monodromy matrix. Using the Poisson bracket relation
for these supertraces with different values of spectral parameter 
(see above) we find that infinite family of IM is involutive, as it was 
mentioned earlier.
\section{Free field representation of superconformal algebra}
In this section we begin to build quantum counterparts of the introduced classical objects. We will start from the quantum Miura transformation, the free 
field representation of the superconformal algebra \cite{SCFT}:
\begin{eqnarray}
-\beta^2T(u)&=&:\phi'^2(u):-(1-\beta^2/2)\phi''(u)+\frac{1}{2}:\xi\xi'(u):+\frac{\epsilon\beta^2}{16}\\ 
\frac{i^{1/2}\beta^2}{\sqrt{2}}G(u)&=&\phi '\xi(u)-(1-\beta^2/2)\xi '(u),
\nonumber
\end{eqnarray}
where
\begin{eqnarray}
&&\phi(u)=iQ+iPu+\sum_n\frac{a_{-n}}{n}e^{inu},\qquad
\xi(u)=i^{-1/2}\sum_n\xi_ne^{-inu},\\
&&[Q,P]=\frac{i}{2}\beta^2 ,\quad 
[a_n,a_m]=\frac{\beta^2}{2}n\delta_{n+m,0},\qquad
\{\xi_n,\xi_m\}=\beta^2\delta_{n+m,0}.\nonumber
\end{eqnarray}
Recall that there are two types of boundary conditions on 
$\xi$: $\xi(u+2\pi)=\pm\xi(u)$. The sign ``+'' corresponds 
to the R sector,the case
when $\xi$ is integer modded, the ``--'' sign corresponds to the NS sector and
$\xi$ is half-integer modded. The variable $\epsilon$ in (13)
is equal to zero
in the R case and equal to 1 in the NS case.\\
One can expand $T(u)$ and $G(u)$ by modes in such a way: $
T(u)=\sum_nL_{-n}e^{inu}-\frac{\hat{c}}{16}$, $G(u)=\sum_nG_{-n}e^{inu}
$,
where  $\hat{c}=5-2(\frac{\beta^2}{2}+\frac{2}{\beta^2})$  and $L_n,G_m$ 
generate the superconformal algebra:
\begin{eqnarray}
[L_n,L_m]&=&(n-m)L_{n+m}+\frac{\hat{c}}{8}(n^3-n)\delta_{n,-m}, \quad
\lbrack L_n,G_m\rbrack=(\frac{n}{2}-m)G_{m+n}\nonumber\\
\lbrack G_n,G_m\rbrack&=&2L_{n+m}+\delta_{n,-m}\frac{\hat{c}}{2}(n^2-1/4).
\end{eqnarray}
In the classical limit  $c\to -\infty$ (the same is $\beta^2\to 0$) 
the following substitution:
$
T(u)\to-\frac{\hat{c}}{4}U(u)$, 
$G(u)\to-\frac{\hat{c}}{2\sqrt{2i}}\alpha(u)$,
$[,]\to \frac{4\pi}{i\hat{c}}\{,\}
$
reduce the above algebra to the Poisson bracket algebra of 
SUSY N=1 KdV theory.\\ 
\hspace*{5mm}Let now $F_p$ be the highest weight 
module over the oscillator algebra of 
$a_n$, $\xi_m$ with the highest weight vector (ground state) $|p\rangle$ 
determined by the 
eigenvalue of $P$ and nilpotency condition of the action of the positive modes:
$
P|p\rangle=p|p\rangle,\quad 
a_n|p\rangle=0, \quad \xi_m|p\rangle=0$ where $n,m > 0$.
In the case of the R sector the highest weight becomes doubly degenerate
due to the presence of zero mode $\xi_0$. So, there are two ground states
$|p,+\rangle$ and $|p,-\rangle$: $|p,+\rangle = \xi_0|p,-\rangle$.
Using the above free field representation of the superconformal algebra
one can obtain that for generic $\hat{c}$ and $p$, $F_p$ is isomorphic to the 
super-Virasoro module with the highest weight vector $|p\rangle$:
$
L_0|p\rangle=\Delta_{NS}|p\rangle,$ where $\Delta_{NS}=
(p/\beta)^2 + (\hat{c}-1)/16$ in the NS sector and module with two highest weight vectors in the Ramond case:
$
L_0|p,\pm\rangle=\Delta_{R}|p,\pm\rangle,\quad\Delta_{R}=
(p/\beta)^2 + \hat{c}/16,\quad
|p,+\rangle=(\beta^2/\sqrt{2}p)G_0|p,-\rangle.
$
The space $F_p$, now considered as super-Virasoro module, splits 
into the sum of finite-dimensional subspaces, determined by the 
value of $L_0$: $
F_p=\oplus^{\infty}_{k=0}F_p^{(k)}$, $L_0 F_p^{(k)}=(\Delta + k) F_p^{(k)}$.
The quantum versions of local integrals of motion should act invariantly on 
the subspaces $F_p^{(k)}$. Thus, the diagonalization
of IM reduces (in a given subspace $ F_p^{(k)}$) to the finite purely 
algebraic problem,
which however rapidly become rather complicated for large $k$. It should 
be noted also that in the case of the 
Ramond sector supersymmetry generator $G_0$ commute with 
IM, so IM act in $|p,+\rangle$ and $|p,-\rangle$ independently, 
without mixing of these
two ground states (unlike the super-KdV case \cite{kulzeit}).
  \section {Quantum monodromy matrix and RTT-relation}
In this part of the work we will consider the quantum $C_q(2)^{(2)}$ 
R-matrix and show that the vertex operator representation of the lower Borel 
subalgebra of $C_q(2)^{(2)}$ allows to represent this R-matrix in the
P-exponent like form which in the classical limit coincide with the
auxiliary L-operator.
$C_q(2)^{(2)}$ is a quantum superalgebra with the following commutation 
relations \cite{tolstkhor}:
\begin{eqnarray}
&&[h_{\alpha_0},h_{\alpha_1}]=0,\quad [h_{\alpha_0},e_{\pm\alpha_1}]= 
\mp e_{\pm\alpha_1},\quad [h_{\alpha_1},e_{\pm\alpha_0}]= 
\mp e_{\pm\alpha_0},\\
&&[h_{\alpha_i},e_{\pm\alpha_i}]=\pm e_{\pm\alpha_i}\quad (i=0,1),\quad
[e_{\pm\alpha_i}, e_{\mp\alpha_j}]=\delta_{i,j}[h_{\alpha_i}]\quad (i,j=0,1),
\nonumber\\
&&[e_{\pm\alpha_1},[e_{\pm\alpha_1},[e_{\pm\alpha_1},
e_{\pm\alpha_0}]_{q}]_{q}]_{q}=0,\quad 
[[[e_{\pm\alpha_1},e_{\pm\alpha_0}]_{q},e_{\pm\alpha_0}]_{q},
e_{\pm\alpha_0}]_{q}=0,
\nonumber
\end{eqnarray}
where $[x]=\frac{q^x-q^{-x}}{q-q^{-1}}$, $p(h_{\alpha_{0,1}})=0$, 
$p(e_{\pm \alpha_{0,1}})=1$ and 
q-supercommutator is defined in the following way:
$[e_{\gamma},e_{\gamma'}]_{q}\equiv e_{\gamma}e_{\gamma'} - 
(-1)^{p(e_{\gamma})p(e_{\gamma'})}
q^{(\gamma,\gamma')}e_{\gamma'}e_{\gamma}$, $q=e^{i\pi\frac{\beta^2}{2}}$.
The corresponding coproducts are:
\begin{eqnarray}
&&\Delta(h_{\alpha_j})=h_{\alpha_j}\otimes 1 + 1\otimes h_{\alpha_j},
\quad
\Delta(e_{\alpha_j})=e_{\alpha_j}\otimes q^{h_{\alpha_j}}+1\otimes 
e_{\alpha_j},\\
&&\Delta(e_{-\alpha_j})=e_{-\alpha_j}
\otimes 1+q^{-h_{\alpha_j}} \otimes e_{-\alpha_j}.\nonumber
\end{eqnarray}
The associated R-matrix can be expressed in such a way \cite{tolstkhor}:
$R=KR_{+}R_{0}R_{-}$, where $K=q^{h_{\alpha}\otimes h_{\alpha}}$, 
$R_{+}=\prod_{n\ge 0}^{\to}R_{n\delta+\alpha}$, 
$R_{-}=\prod_{n\ge 1}^{\gets}R_{n\delta-\alpha}$,
$R_{0}=\exp((q-q^{-1})\sum_{n>0}d(n)e_{n\delta}\otimes e_{-n\delta}).$
Here $R_{\gamma}=\exp_{(-q^{-1})}(A(\gamma)(q-q^{-1})(e_{\gamma}
\otimes e_{-\gamma}))$ and coefficients $A$ and $d$ are defined as follows:
$A(\gamma)=\{(-1)^n$ if $\gamma=n\delta+\alpha; (-1)^{n-1}$ if 
$\gamma=n\delta-\alpha\}$, $d(n)=\frac{n(q-q^{-1})}{q^n-q^{-n}}$.
The generators $e_{n\delta}$, $e_{n\delta\pm\alpha}$ are defined via the 
q-commutators of Chevalley generators. In the following we will need  
the expressions only for simplest ones: 
$e_{\delta}=[e_{\alpha_0}, e_{\alpha_1}]_{q^{-1}}$ and 
$e_{-\delta}=[e_{-\alpha_1},e_{-\alpha_0}]_q$. The elements 
$e_{n\delta\pm\alpha}$ are expressed as multiple commutators of 
$e_{\delta}$ with corresponding Chevalley generators, $e_{n\delta}$ ones 
have more complicated form \cite{tolstkhor}.\\    
\hspace*{5mm}Let's introduce the reduced R-matrix $\bar{R}\equiv K^{-1}R$.
Using all previous information one can write $\bar{R}$ as
$\bar{R}(\bar{e}_{\alpha_i},\bar{e}_{-\alpha_i})$, where
$\bar{e}_{\alpha_i}=e_{\alpha_i}\otimes 1$ and
$\bar{e}_{-\alpha_i}=1\otimes e_{-\alpha_i}$, because it is
represented as power series of these elements.
After this necessary background we will 
introduce vertex operators and using the fact that they
satisfy the Serre relations of the lower Borel subalgebra of $C_q(2)^{(2)}$  
we will prove that the reduced R-matrix, represented by the vertex operators 
has the properties of the P-exponent.
So, the vertex operators are:
$
V_1=\frac{1}{q^{-1}-q}\int \d\theta \int^{u_{1}}_{u_{2}} \d u :e^{-\Phi}:,
$ 
$
V_0=\frac{1}{q^{-1}-q}\int \d\theta \int^{u_{1}}_{u_{2}} \d u :e^{\Phi}:,
$
where $2\pi \ge u_1 \ge\ u_2\ge 0$, $\Phi=\phi(u) - 
\frac{i} {\sqrt{2}}\theta\xi(u)$ is a superfield and normal ordering 
here means that
$:e^{\pm\phi(u)}:=
\exp\Big(\pm\sum_{n=1}^{\infty}\frac{a_{-n}}{n}e^{inu}\Big)
\exp\Big(\pm i(Q+Pu)\Big)\exp\Big(\mp\sum_{n=1}^{\infty}\frac{a_{n}}{n}e^{-inu}
\Big)$. 
One can show via the standard contour technique that these 
operators satisfy the
same commutation relations as $e_{\alpha_1}$, $e_{\alpha_0}$ correspondingly.\\
\hspace*{5mm}Then, following \cite{4} one can show 
(using the fundamental property of the universal R-matrix: 
$(I\otimes\Delta)R=R^{13}R^{12}$) 
that the reduced R-matrix has the following 
property: 
\begin{equation}
\bar{R}(\bar{e}_{\alpha_i},e'_{-\alpha_i}+e''_{-\alpha_i})=
\bar{R}(\bar{e}_{\alpha_i},e'_{-\alpha_i})\bar{R}(\bar{e}_{\alpha_i},
e''_{-\alpha_i}),
\end{equation} 
where $e'_{-\alpha_i}=1\otimes q^{-h_{\alpha_i}}\otimes 
e_{-\alpha_i}$, 
$e''_{-\alpha_i}=1\otimes e_{-\alpha_i}\otimes 1$,
$\bar{e}_{\alpha_i}=e_{\alpha_i} \otimes 1\otimes 1$. 
The commutation relations between them are:
\begin{eqnarray}
e'_{-\alpha_i}\bar{e}_{\alpha_j}&=&-\bar{e}_{\alpha_j}e'_{-\alpha_i},\quad
e''_{-\alpha_i}\bar{e}_{\alpha_j}=-\bar{e}_{\alpha_j}e''_{-\alpha_i},\\
e'_{-\alpha_i}e''_{-\alpha_j}&=&-q^{b_{ij}}e''_{-\alpha_j}e'_{-\alpha_i},
\nonumber
\end{eqnarray}
where $b_{ij}$ is the symmetric matrix with the following elements:
$b_{00}=b_{11}=-b_{01}=1$. Now, denoting $\mathbf{\bar{L}}^{(q)}(u_2, u_1)$ 
the reduced R-matrix with $e_{-\alpha_i}$ represented by $V_{i}$, we find,
using the above property of $\bar{R}$ with $e'_{-\alpha_i}$ replaced by 
appropriate vertex operators: 
$\mathbf{\bar{L}}^{(q)}(u_3, u_1)=\mathbf{\bar{L}}^{(q)}(u_2, u_1)
\mathbf{\bar{L}}^{(q)}(u_3, u_2)$ with $u_1 \ge u_2\ge u_3 $. 
So, $\mathbf{\bar{L}}^{(q)}$ has the property of P-exponent. 
But because of singularities in the operator 
products of vertex operators it can not be written in the usual P-ordered
form. Thus, we propose a new notion, the quantum P-exponent: 
\begin{equation}
\mathbf{\bar{L}}^{(q)}(u_1, u_2)=Pexp^{(q)}\int^{u_1}_{u_2}\d u \int\d \theta
(e_{\alpha_1} :e^{-\Phi}:+e_{\alpha_0}:e^{\Phi}:).
\end{equation}
Introducing new object: $\mathbf{L}^{(q)}\equiv e^{i\pi Ph_{\alpha_1}}
\mathbf{\bar{L}}^{(q)}(0,2\pi)$, which coincides with R-matrix with 
$1\otimes h_{\alpha_1}$ replaced by $2P/\beta^2$ and 
$1\otimes e_{-\alpha_1}$, $1\otimes e_{-\alpha_0}$ replaced by $V_1$ and 
$V_0$ (with integration from 0 to $2\pi$) correspondingly, 
we find that it satisfies the well-known RTT-relation:
\begin{eqnarray}
&&\mathbf{R}(\lambda\mu^{-1})
\Big(\mathbf{L}^{(q)}(\lambda)\otimes \mathbf{I}\Big)\Big(\mathbf{I}
\otimes \mathbf{L}^{(q)}(\mu)\Big)\\
&&=(\mathbf{I}\otimes \mathbf{L}^{(q)}(\mu)\Big)
\Big(\mathbf{L}^{(q)}(\lambda)\otimes \mathbf{I}\Big)\mathbf{R}
(\lambda\mu^{-1}), \nonumber
\end{eqnarray}
where the dependence on $\lambda,\mu$ means that we are considering
$\mathbf{L}^{(q)}$-operators in the evaluation representation of 
$C_{q}(2)^{(2)}$. 
Thus the supertraces of ``shifted'' $\mathbf{L}^{(q)}$-operators, the 
transfer matrices
$\mathbf{t}^{(q)}(\lambda)\equiv str (e^{i\pi Ph_{\alpha_1}}
\mathbf{L}^{(q)}(\lambda))$ 
commute:\\ $[\mathbf{t}^{(q)}(\lambda),\mathbf{t}^{(q)}(\mu)]=0$,
giving the quantum integrability.\\
\hspace*{5mm}Now we will show that in the classical
limit ($q\to1$) the $\mathbf{L}^{(q)}$-operator will give the 
auxiliary $\mathbf{L}$-matrix defined in the Sec. 2.  
We will use the P-exponent property of $\mathbf{\bar{L}}^{(q)}(0, 2\pi)$.
Let's decompose $\mathbf{\bar{L}}^{(q)}(0, 2\pi)$ in the following way:
$
\mathbf{\bar{L}}^{(q)}(0, 2\pi)=
\lim_{N\to\infty}\prod_{m=1}^{N}\mathbf{\bar{L}}^{(q)}(x_{m-1},x_{m})
$, where we divided  the interval $[0,2\pi]$ into infinitesimal 
intervals $[x_m,x_{m+1}]$
with $x_{m+1}-x_m=\epsilon=2\pi/N$.
Let's find the terms that can give contribution of the first order
in $\epsilon$ in $\mathbf{\bar{L}}^{(q)}(x_{m-1},x_{m})$. 
In this analysis we will need the operator product expansion of 
vertex operators:
\begin{eqnarray}
&&\xi(u)\xi(u')=
-\frac{i\beta^2}{(iu-iu')}+\sum_{k=1}^{\infty}c_k(u)(iu-iu')^k,\\
&&:e^{a\phi(u)}::e^{b\phi(u')}:=(iu-iu')^{\frac{ab\beta^2}{2}}
(:e^{(a+b)\phi(u)}:+\sum_{k=1}^{\infty}d_k(u)(iu-iu')^k),\nonumber
\end{eqnarray}
where $c_k(u)$ and $d_k(u)$ are operator-valued functions of $u$.
Now one can see that only two types of terms 
can give the contribution of the order $\epsilon$ in 
$\mathbf{\bar{L}}^{(q)}(x_{m-1},x_{m})$ when $q\to 1$.
The first type consists of operators of the first order in $V_i$ 
and the second type is formed
by the operators, quadratic in $V_i$, which give contribution of the order
$\epsilon^{1\pm\beta^2}$ by virtue of operator product expansion.
Let's look on the terms of the second type in detail.
At first we consider the terms 
appearing from the $R_0$-part of R-matrix, represented by vertex operators:
\begin{eqnarray}
&&\frac{e_{\delta}}{2(q-q^{-1})}\Bigg(\int^{x_m}_{x_{m-1}}\d u_1 :e^{-\phi}:
\xi(u_1-i0)\int^{x_m}_{x_{m-1}}\d u_2 :e^{\phi}:\xi(u_2+i0)+\\
&&q^{-1}\int^{x_m}_{x_{m-1}}\d u_2 :e^{\phi}:
\xi(u_2-i0)\int^{x_m}_{x_{m-1}}\d u_1 :e^{-\phi}:\xi(u_1+i0)\Bigg)\nonumber
\end{eqnarray}
Neglecting the terms, which give rise to $O(\epsilon^2)$ contribution, 
we obtain, using the operator products of vertex operators:
\begin{eqnarray}
&&\frac{e_{\delta}}{2(q-q^{-1})}\Bigg(\int^{x_m}_{x_{m-1}}\d u_1 
\int^{x_m}_{x_{m-1}}\d u_2 \frac{-i\beta^2}{(i(u_1-u_2-i0))^{\frac{\beta^2}{2}
+1}}+\\
&&q^{-1}\int^{x_m}_{x_{m-1}}\d u_2 \int^{x_m}_{x_{m-1}}\d u_1 
\frac{-i\beta^2}{(i(u_2-u_1-i0))^{\frac{\beta^2}{2}+1}}\Bigg)\nonumber
\end{eqnarray}
In the $\beta^2\to 0$ limit we get:
$
\frac{[e_{\alpha_1},e_{\alpha_0}]}{2i\pi}
\int^{x_m}_{x_{m-1}}\d u_1 
\int^{x_m}_{x_{m-1}}\d u_2 \big(\frac{1}{u_1-u_2+i0}
-\frac{1}{u_1-u_2-i0}\big)\nonumber
$
which, by the well known formula: $\frac{1}{x+i0}=P\frac{1}{x} -i\pi\delta(x)$
gives: $-\int^{x_m}_{x_{m-1}}\d u [e_{\alpha_1}, e_{\alpha_0}]$.
Another terms arise from the $R_{+}$ and $R_{-}$ parts of R-matrix 
and are very similar to each other:
\begin{eqnarray}
\frac{e_{\alpha_0}^2}{2(2)_{(-q^{-1})}}\int^{x_m}_{x_{m-1}}
\d u_1 :e^{\phi}:\xi(u_1-i0)
\int^{x_m}_{x_{m-1}}\d u_2 :e^{\phi}:\xi(u_2+i0),\\
\frac{e_{\alpha_1}^2}{2(2)_{(-q^{-1})}}\int^{x_m}_{x_{m-1}}
\d u_1 :e^{-\phi}:\xi(u_1-i0)
\int^{x_m}_{x_{m-1}}\d u_2 :e^{-\phi}:\xi(u_2+i0).\nonumber
\end{eqnarray}
The integrals can be reduced to the ordered ones: 
\begin{eqnarray}
\frac{e_{\alpha_0}^2}{2}\int^{x_m}_{x_{m-1}}\d u_1 :e^{\phi}:\xi(u_1)
\int^{u_1}_{x_{m-1}}\d u_2 :e^{\phi}:\xi(u_2),\\
\frac{e_{\alpha_1}^2}{2}\int^{x_m}_{x_{m-1}}\d u_1 :e^{-\phi}:\xi(u_1)
\int^{u_1}_{x_{m-1}}\d u_2 :e^{-\phi}:\xi(u_2).\nonumber
\end{eqnarray}
Following \cite{kulzeit} we find that their contribution (of order $\epsilon$) in 
the classical limit is:
\begin{eqnarray}
-e_{\alpha_0}^2\int^{x_m}_{x_{m-1}}\d u e^{2\phi(u)},\quad
-e_{\alpha_1}^2\int^{x_m}_{x_{m-1}}\d u e^{-2\phi(u)}.
\end{eqnarray} 
Gathering now all the terms of order $\epsilon$ we find:
\begin{eqnarray}
&&\mathbf{\bar{L}}^{(1)}(x_{m-1},x_{m})=1+\int^{x_m}_{x_{m-1}}\d u 
(\frac{i}{\sqrt{2}}
\xi(u)e^{-\phi(u)}e_{\alpha_1}-\\
&&\frac{i}{\sqrt{2}}
\xi(u)e^{\phi(u)}e_{\alpha_0}
-e^2_{\alpha_1}e^{-2\phi(u)}-
e^2_{\alpha_1}e^{2\phi(u)}-[e_{\alpha_1},e_{\alpha_0}])+O(\epsilon^2)
\nonumber
\end{eqnarray}
and collecting all $\mathbf{\bar{L}}^{(1)}(x_{m-1},x_{m})$ we find
that $\mathbf{\bar{L}}^{(1)}$=$e^{-i\pi p h_{\alpha_1}}\mathbf{L}$. Therefore 
$\mathbf{L}^{(1)}$=$\mathbf{L}$.
\section{Final remarks}
The obtained RTT-relation (21) and 
corresponding quantum integrability condition 
give us possibility to consider the model possessing the associated IM from a 
point of view of QISM \cite{leshouches}, 
\cite{kulsklyan}.
In our case the class of such systems is very wide. It contains all the superconformal (minimal) models and some of their 
perturbations (though in the unperturbed case SCFT usually provides 
easier methods). The perturbations that do not 
break the integrability (commuting with IM) and therefore appropriate 
for the QISM 
scheme are (following the arguments of \cite {1},\cite{fioravanti}) 
$\phi_{(1,3)}$ and $\phi_{(3,1)}$ operators, corresponding to 
the vertex operators $\int_{0}^{2\pi}\d u\int\d\theta e^{\pm\Phi}$.
The topic of special interest is the SCFT with perturbation on the boundary  
related to the D-brane theory \cite{moore}, \cite{vitchev}.\\
\hspace*{5mm}In order to find the eigenvalues  of the transfer matrices 
(they are our main object of study, because their expansion in $\lambda$ 
gives the quantum IM) one can follow two routes. The first one, introduced in
\cite{1}, is based on the so-called ``fusion relations'' \cite{kulsklyan} 
for the transfer 
matrices. In the case when $q$ is a root of unity ($\beta^2$ and $c$ are 
rational), the 
system of fusion relations becomes finite and reduces to the Thermodynamic 
Bethe
Ansatz equations \cite{TBA}.\\
\hspace*{5mm} Another approach is the Baxter Q-operator method, which could be 
applied for all values of $\beta^2$ (therefore for all values of the central 
charge)\cite{2}. The construction of the Q-operator and 
fusion relations will be given in the paper under preparation \cite{prepar}.

\section*{Acknowledgments}
We are grateful to F.A. Smirnov for useful discussions.
The work was supported by the Dynasty Foundation (AMZ) and 
RFBR grant 03-01-00593 (PPK).


\begin{thebibliography}{9}
\bibitem{gervais}A. Bilal, J.-L. Gervais, Phys. Lett. B 211 (1988) 95. 

\bibitem{kulzeit}P.P. Kulish, A.M. Zeitlin, Phys. Lett. B 581 (2004) 125.  

\bibitem{sKdV1}B.A. Kupershmidt, Phys.  Lett. A 102 (1984) 213.   

\bibitem{sKdV2}P.P. Kulish, Zap. Nauchn. Sem. LOMI 155 (1986) 142. 

\bibitem{sKdV3}P.P. Kulish, A.M. Zeitlin, 
Zap. Nauchn. Sem. POMI 291 (2002) 185. 

\bibitem{mathieu} P. Mathieu, Phys. Lett. B 203 (1988) 287.

\bibitem{inami}T. Inami, H. Kanno, Comm. Math. Phys. 136 (1991) 519.

\bibitem{tolstkhor}S.M. Khoroshkin, J. Lukierski, 
V.N. Tolstoy, math.QA/0005145.

\bibitem{leshouches}L.D. Faddeev, in: 
Quantum symmetries/Symmetries quantiques, eds. 
A.Connes et al. (North-Holland 1998) p. 149.

\bibitem{kulsklyan}P.P. Kulish, E.K. Sklyanin, 
Lect. Notes Phys. 151 (1982) 61.

\bibitem{qiu}Z. Qiu, Nucl. Phys. B 270 (1986) 205.

\bibitem{moore}G. Moore, hep-th/0304018.

\bibitem{vitchev}E.S. Vitchev, hep-th/0404195.

\bibitem{1} V.V. Bazhanov, S.L. Lukyanov, A.B. Zamolodchikov,
 Comm. Math. Phys. 177 (1996) 381.

\bibitem{fadd}L.D. Faddeev, L.A. Takhtajan, Hamiltonian Method in the Theory
of Solitons (Springer 1987). 

\bibitem{shadrikov} V.V. Bazhanov, A.G. Shadrikov, Theor. Math. Phys. 73 (1988) 1302.

\bibitem{SCFT}M.A. Bershadsky, V.G. Knizhnik, M.G. Teitelman,
 Phys. Lett. B 151 (1985) 31.

\bibitem{4}V.V. Bazhanov, A.N. Hibberd, S.M. Khoroshkin, Nucl. Phys. B 622 
(2002) 475.

\bibitem{fioravanti}D. Fioravanti, F. Ravanini, M. Stanishkov,
 Phys. Lett. B 367 (1996) 113.

\bibitem{TBA}Al.B. Zamolodchikov, Phys. Lett. B 253 (1991) 391. 

\bibitem{2} V.V. Bazhanov, S.L. Lukyanov, A.B. Zamolodchikov,
 Comm. Math. Phys. 200 (1997) 247.

\bibitem{prepar}
 P.P. Kulish ,  A.M. Zeitlin, in preparation.

\end{thebibliography}
\end{document}